\documentclass[aps,prl,reprint]{revtex4-1}
\usepackage{graphicx,color,dcolumn,bm,amsfonts,amsmath,amssymb}
\usepackage{mathptmx,longtable,float,setspace,soul} 
\usepackage{xcolor}
\usepackage{multirow}

\begin{document}

\title{Why Fe$_{3}$GaTe$_{2}$ has higher Curie temperature than Fe$_{3}$GeTe$_{2}$?}

\author{Bomin Kim$^1$}
\author{Tumentsereg Ochirkhuyag$^{2}$}
\author{Dorj Odkhuu$^2$}
\email{odkhuu@inu.ac.kr}
\author{S. H. Rhim$^1$}
\email{sonny@ulsan.ac.kr}
\affiliation{$^1$Department of Physics, University of Ulsan, Ulsan 44610, Republic of Korea\\
$^2$Department of Physics, Incheon National University, Incheon 22012, Republic of Korea}

\date{\today}

\begin{abstract} 
Physics of Fe$_3$GaTe$_2$ having higher Curie temperature ($T_C$) than Fe$_3$GeTe$_2$ is explored theoretically in the framework of magnetic exchange interactions. 
Fe$_3$GaTe$_2$ and Fe$_3$GeTe$_2$ are isostructural, with Fe$_3$GaTe$_2$ 
having one less valence electron and smaller nearest-neighbor exchange coefficients ($J_1$ and $J_2$), 
challenging the conventional notion that larger $J_1$ or $J_2$ leads to a higher $T_C$.
We show that higher order exchange coefficients, $J_3$ or higher, of Fe$_3$GaTe$_2$ are positive whereas those of Fe$_3$GeTe$_2$ are negative. As a consequence, total sum of all possible exchange coefficients in Fe$_3$GaTe$_2$ are larger than Fe$_3$GeTe$_2$, which accounts for higher $T_C$. 
To validate these findings, $T_C$ are computed using both mean-field theory and Monte Carlo simulation.  
Indeed, higher-order exchange interactions, when properly accounting for the number of neighbors, 
confirm the higher $T_C$ of Fe$_3$GaTe$_2$.

\end{abstract}

\maketitle


\emph{Introduction}$-$
Two-dimensional van der Waals magnets offer unique opportunities for investigating fundamental magnetic models such as Ising, XY, and Heisenberg systems, 
providing a fertile ground for examining classic phenomena, 
including the Onsager solution \cite{onsager1944crystal}, 
the Berezinskii-Kosterlitz-Thouless transition \cite{kosterlitz1973ordering}, 
and various forms of magnetic anisotropy \cite{gibertini2019magnetic, burch2018magnetism}.

Extensive research has been conducted \cite{burch2018magnetism, gibertini2019magnetic,huang2017layer, gong2017discovery} to achieve Curie temperatures ($T_C$) above room temperature.
Among these, Fe$_{3}$GaTe$_{2}$ has attracted particular attention with $T_C$=380 K surpassing room temperature \cite{li2023tuning,jin2023room,wang2023sign}.
Fe$_{3}$GaTe$_{2}$ is isostructural to Fe$_{3}$GeTe$_{2}$, replacing Ge by Ga with one less valence electron. 
The observation that the Curie temperature ($T_C$) of Fe$_3$GaTe$_2$ exceeds that of Fe$_3$GeTe$_2$ 
by over 100 K, despite differing by only one electron, warrants systematic investigation \cite{zhang2022above,deiseroth2006fe3gete2}.

From previous study \cite{burch2018magnetism}, Fe$_{3}$GaTe$_{2}$ and Fe$_{3}$GeTe$_{2}$ are presumably itinerant ferromagnet, whose spin Hamiltonian is conveniently expressed as sum of the Heisenberg exchange and the magnetic anisotropy,
$H = -\sum_{i\neq j}J_{ij}\textbf{S}_i\cdot \textbf{S}_j-K_u\sum_{i}(\textbf{S}_i\cdot \textbf{e})^2$.
$J_{ij}$ is the magnetic exchange coefficient of two spins at atomic site $i$ and $j$; 
$K_u$ is the magnetic anisotropy, where $\textbf{e}$ is the unit vector along the magnetic easy axis.
Noticeable $T_C$ difference of two materials is intensively studied \cite{marfoua2024large, wu2024spectral}
in the framework of aforementioned spin Hamiltonian, more specifically in terms of magnetic exchange coefficients.
It was attributed that larger nearest neighbor exchange coefficient, $J_1$, of Fe$_3$GaTe$_2$
is responsible for higher $T_C$ \cite{lee2023electronic}.
In other study, on the other hand, it was suggested that the magnetic exchange coefficients of third nearest neighbor, $J_3$ and $J'_3$, are the main factor of higher $T_C$ of Fe$_3$GaTe$_2$ \cite{ruiz2024origin}. 
In this letter, after examining both the magneto-crystalline anisotropy and the magnetic exchange coefficients, we show that higher-order magnetic exchange coefficients are indispensable for higher $T_C$ of Fe$_3$GaTe$_2$. Moreover, the magnetic exchange coefficients with fully accounting number of neighbors are necessary in $T_C$ estimation.

\begin{figure}[t]
    \centering
    \includegraphics[width=1\columnwidth]{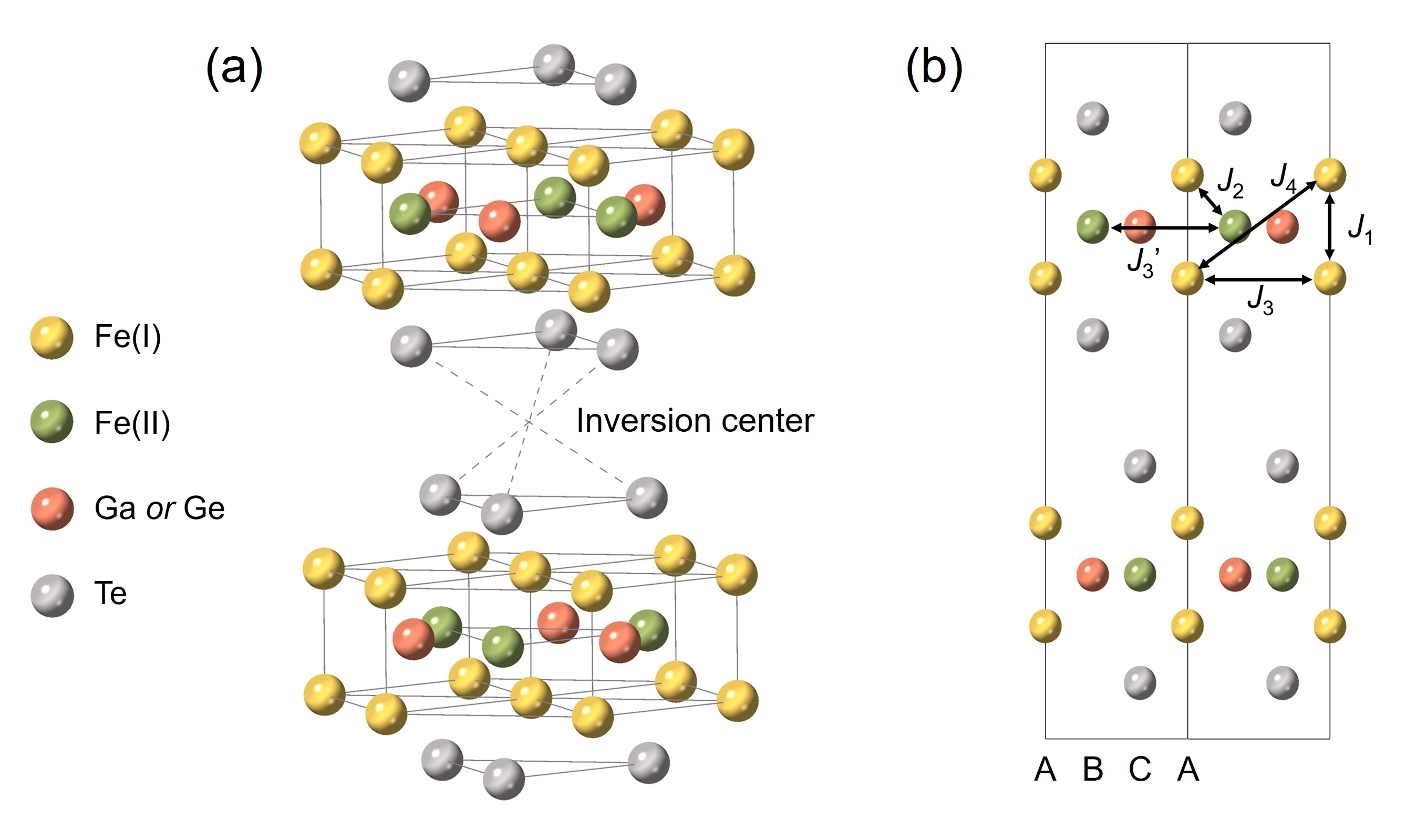}
    \caption{(a) Crystal structure of Fe$_{3}$GaTe$_{2}$ and Fe$_{3}$GeTe$_{2}$. 
    (b) The exchange parameters (from $J_{1}$ to $J_{4}$) are denoted in the structure. $J_{1}$, $J_{3}$, and $J_{4}$ indicate the interaction of Fe(I)-Fe(I); $J_{2}$ and $J'_{3}$ for Fe(I)-Fe(II) and Fe(II)-Fe(II), respectively. }
    \label{fig:structure}
\end{figure}

Fig. \ref{fig:structure} presents structure of Fe$_{3}$GaTe$_2$ and Fe$_{3}$GeTe$_2$ with different center atom, Ga or Ge. Both Fe$_{3}$GaTe$_2$ and Fe$_{3}$GeTe$_2$ crystallize in a hexagonal structure with space group P6$_{3}/mmc$ (No. 194). The structure consists of one bilayer or two monolayer units, which are connected by the inversion symmetry. The monolayer unit contain quintuple sublayers. Two symmetrically distinct Fe atoms are distinguished as Fe(I) and Fe(II). The Fe(II)-Ga (or Ge) layer sandwiched by Fe(I) and Te. 
Exchange interactions from $J_{1}$ to $J_{4}$ are shown in Fig. \ref{fig:structure} (b), where the intra-layer interactions are identical in both layers. A detailed discussion will be provided later.

\emph{Computational details}$-$Density-functional calculations are carried out using the Vienna $ab$ $initio$ simulation package (VASP) \cite{kresse1996efficient}. For Brillouin zone summation, 15$\times$15$\times$3 $k$ mesh in $\Gamma$-centered scheme is used. Energy cutoff for planewave expansion is 500 eV.
The generalized gradient approximation (GGA) with Perdew, Burke, and Ernzerhof (PBE) parametrization \cite{perdew1996generalized} is used for the exchange-correlation potential, where the interlayer van der Waals interactions are treated by the DFT-D3 method \cite{grimme2010consistent}.
The atomic positions are optimized with the force criterion of 1 meV/$\AA$. Additional self-consistent calculations are performed using the OpenMX package \cite{PhysRevB.67.155108, PhysRevB.69.195113} which is based on the LCPAO (linear combination of pseudo-atomic orbitals). The atomic basis sets for Fe, Ga, Ge and Te are H-$s3p2d2f1, s2p2d2, s2p2d1$, and $s2p2d2f1$, respectively, with energy cutoff of 300 Ry and cutoff radii of 6.0 (Fe) and 7.0 (Ga, Ge and Te) a.u. (atomic unit). In the framework of the magnetic force theorem (MFT) \cite{liechtenstein1987local}, the exchange coefficients are calculated using the Heisenberg model through implementation of the J$_{x}$ package \cite{yoon2018jx, yoon2020jx, CiteJx}. From the exchange coefficients, $T_C$ are carried out using Monte Carlo simulation in the VAMPIRE package \cite{metropolis1953equation, evans2014atomistic, vampire, CiteVampire}.

\begin{figure}[t]
    \centering
    \includegraphics[width=1\columnwidth]{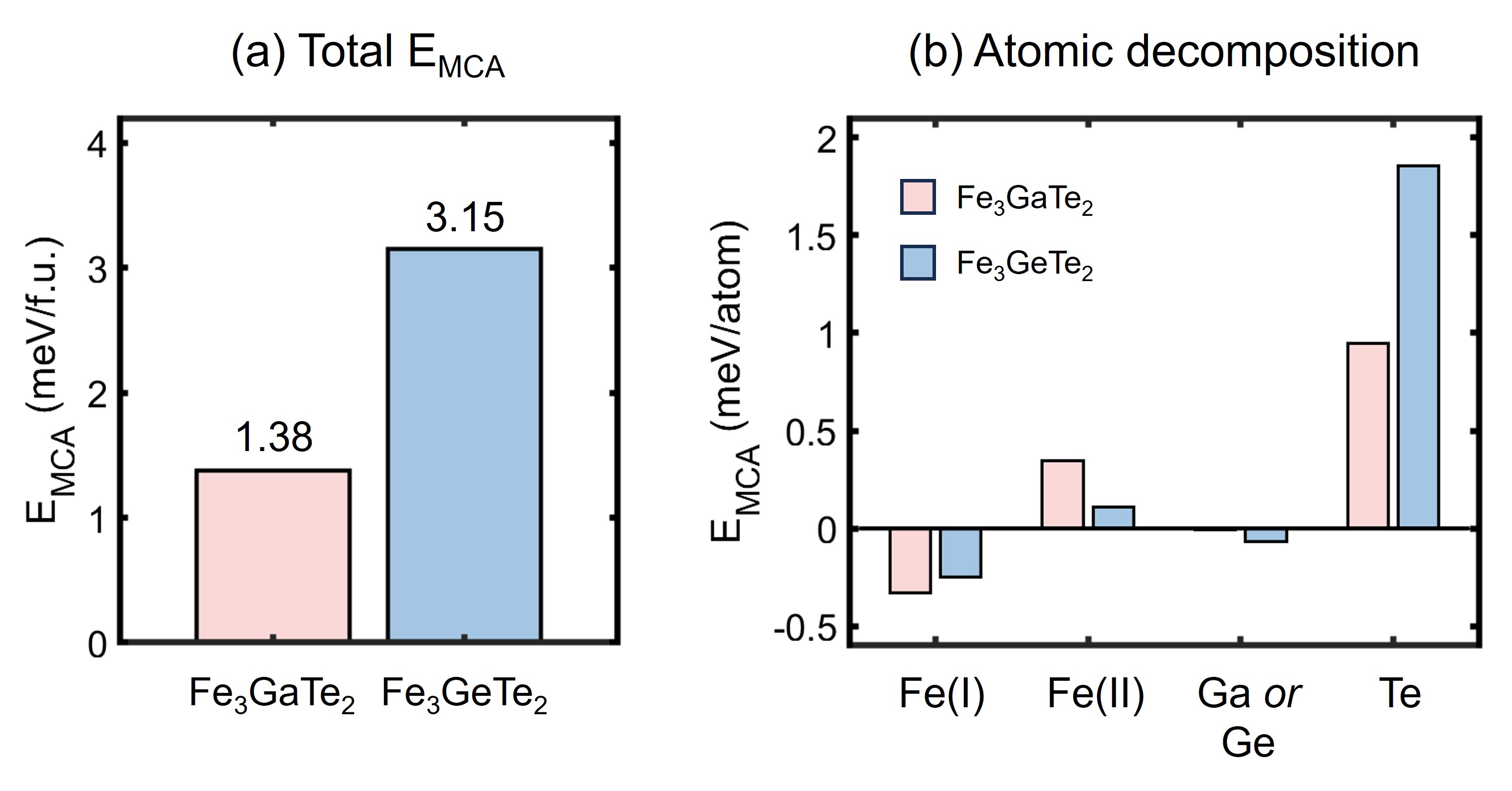}
    \caption{(a) Total $E_{\text{MCA}}$ (in meV/f.u.) of Fe$_{3}$GaTe$_{2}$ (1.38) and Fe$_{3}$GeTe$_{2}$ (3.15). 
    (b) Atomic decomposition of $E_{\text{MCA}}$ (in meV/atom) into Fe(I), Fe(II), Ga/Ge, and Te. Red and blue denote Fe$_{3}$GaTe$_{2}$ and Fe$_{3}$GeTe$_{2}$, respectively.}
    \label{fig:MCA}
\end{figure}

\begin{figure*}[htb]
    \centering
    \includegraphics[width=1\textwidth]{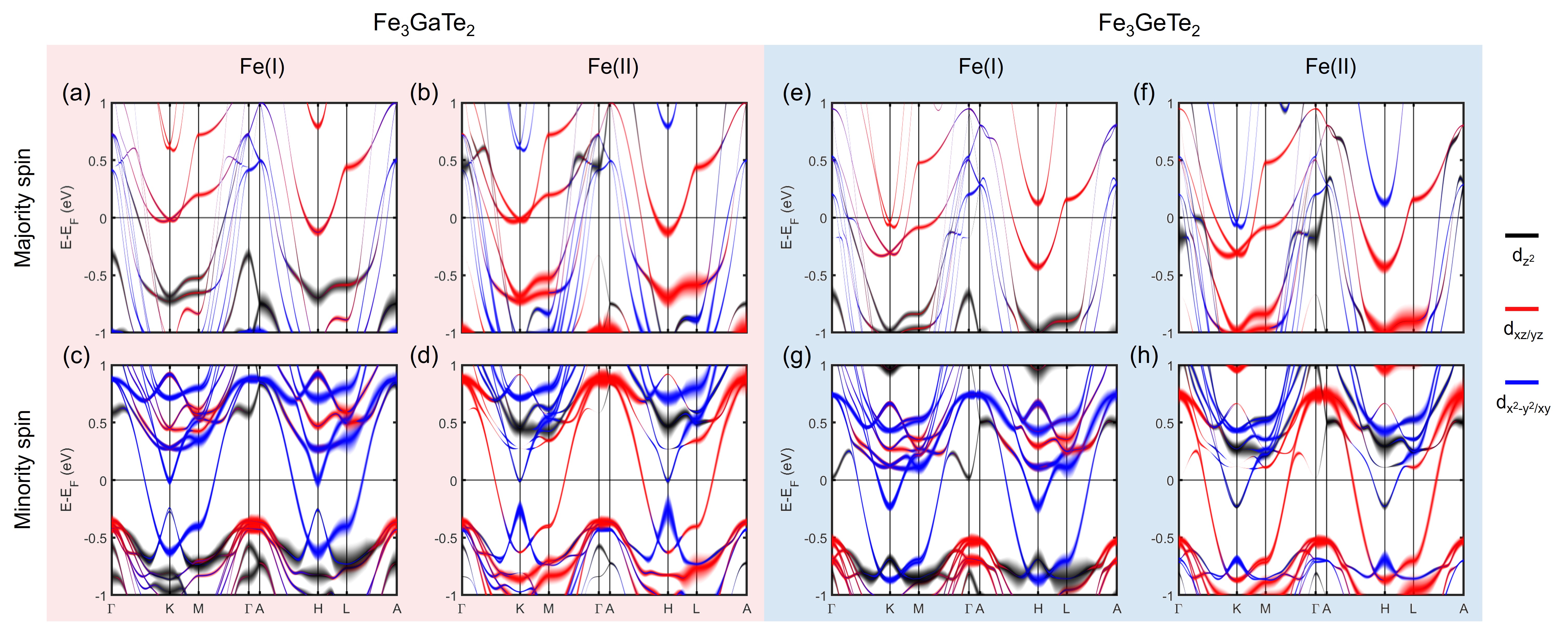}
    \caption{Band structures in $d$ orbital projection. Fe$_{3}$GaTe$_{2}$ and Fe$_{3}$GeTe$_{2}$ in red box and blue box, respectively. In each box, left and right panels for Fe(I) and Fe(II), respectively. Upper (lower) panels for the majority (minority) spin channel. $d$ orbital decomposition into $d_{z^{2}}$ (black line), $d_{xz/yz}$ (red line) and $d_{x^{2}-y^{2}/xy}$ (blue line) according to the irreducible representation of hexagonal symmetry.}
    \label{fig:band}
\end{figure*}


\emph{Magneto-crystalline anisotropy}$-$
Magneto-crystalline anisotropy (MCA) energy, $E_{\text{MCA}}$ is calculated from the total energy difference, $E_{\text{MCA}} = E(\parallel) - E(\perp)$, where $\parallel$ and $\perp$ indicate in-plane and perpendicular magnetization, respectively, with spin-orbit coupling (SOC) included.
Fig. \ref{fig:MCA} (a) shows total $E_{\text{MCA}}$ for Fe$_{3}$GaTe$_{2}$ and  Fe$_{3}$GeTe$_{2}$ in red and blue bar, respectively. $E_{\text{MCA}}=$ 1.38 meV/f.u. of Fe$_{3}$GaTe$_{2}$ is smaller than $E_{\text{MCA}}=$ 3.15 meV/f.u. of Fe$_{3}$GeTe$_{2}$,
which agrees well with previous studies \cite{zhang2022above, li2023tremendous, zhuang2016strong, wang2020modifications, kim2024strain}.
As presented in Fig. \ref{fig:MCA} (b) for atomic-decomposed MCA,
in both Fe$_{3}$GaTe$_{2}$ and Fe$_{3}$GeTe$_{2}$, 
contributions of two Fe sites exhibit different signs: 
$E_{\text{MCA}}[\text{Fe(I)}]<$ 0 and $E_{\text{MCA}}[\textrm{Fe(II)}]>$ 0, 
while $|E_{\text{MCA}}[\textrm{Fe(I)}]|$ and $|E_{\text{MCA}}[\textrm{Fe(II)}]|$ are larger in Fe$_{3}$GaTe$_{2}$.
 
For MCA analysis, we focus on Fe sites although contributions of Te are much larger. Te contributions, from band analysis, are due to occupation change
of $p_x$ and $p_y$, which are strongly hybridized with Fe $d$ orbitals.
Hence, without loss of generality Fe $d$ orbital based analysis is sufficient.
We provide Te $p$ orbital based analysis in Supplementary Material \cite{SMTe}.
Fig. \ref{fig:band} shows band structures of Fe(I) and Fe(II), where $d$ orbitals are decomposed into the irreducible representations \cite{Sakurai, dresselhaus2007group} $d_{z^2}$, $d_{xz/yz}$, and $d_{x^2-y^2/xy}$ or $m=0$, $\pm1$, and $\pm2$ in terms of magnetic quantum number $m$, respectively.
In Fig.\ref{fig:band}, red (blue) box denotes Fe$_{3}$GaTe$_{2}$ (Fe$_{3}$GeTe$_{2}$); left and right panel of each box is for Fe(I) and Fe(II), respectively. Upper (lower) panels are for the majority (minority) spin channel, denoted hereafter as $\uparrow$ ($\downarrow$). 
In rigid-band picture, Fe$_3$GaTe$_2$ bands are downward shift of $E_F$ of Fe$_3$GeTe$_2$ bands due to one less valence electron.

In MCA analysis,
the framework of the second-order perturbation theory is employed \cite{wang1993first, odkhuu2016FeMgO,kim2024strain, ho2022first, PhysRevB.101.214436},
\begin{equation}
    \emph{E}_{\text{MCA}}=\xi^2\sum_{\sigma,\sigma',o,u} \frac{|\langle o,\sigma|L_z|u,\sigma'\rangle|^2-|\langle o,\sigma|L_x|u,\sigma'\rangle|^2}{E_{u,\sigma}-E_{o,\sigma'}}.
    \label{eq:perturbation}
\end{equation}
$\xi$ is the strength of spin-orbit coupling; $o$ ($u$) stands for occupied (unoccupied) state; $\sigma$ and $\sigma'$ denote spin states; $L_{z}$ ($L_{x}$) is the orbital angular momentum operator for $z$ ($x$) component. 
In the following, MCA analysis is presented based on $d$ manifold of Fe(I) and Fe(II)
by comparing Fe$_3$GaTe$_2$ and Fe$_3$GeTe$_2$ band structure.  Fe$_3$GaTe$_2$ is discussed first and then relative change in Fe$_3$GeTe$_2$ follows.

For Fe(I) in Fe$_{3}$GaTe$_{2}$, $E_{\text{MCA}}<0$ arises through the spin-flip channel, $\langle d_{xz/yz},\uparrow (\downarrow)|L_z|d_{xz/yz},\downarrow (\uparrow)\rangle$, where bra (ket) denotes occupied (unoccupied) state. Spin states with parenthesis, $\uparrow$ ($\downarrow$) and $\downarrow$ ($\uparrow$), indicate states in both $\uparrow$ and $\downarrow$ states.
As mentioned earlier, in a simplistic picture, Fe$_3$GaTe$_2$ bands are downward shift of $E_F$ of Fe$_3$GeTe$_2$ bands.
For Fe$_3$GeTe$_2$, as $d_{xz/yz}$ bands become occupied near $KM$ and at $H$ in $\uparrow$ channel, $E_{\text{MCA}}>0$ contributions from $\langle d_{xz/yz},\uparrow|L_z|d_{xz/yz},\uparrow \rangle$ increase. As a consequence, $|E_{\text{MCA}} [\textrm{Fe(I)}]|$ reduces in Fe$_{3}$GeTe$_{2}$.

For Fe(II) in Fe$_{3}$GaTe$_{2}$, $E_{\text{MCA}}>0$ comes from the same spin channel, $\langle d_{xz/yz},\uparrow (\downarrow)|L_z|d_{xz/yz},\uparrow (\downarrow) \rangle$, except near $K$ and $H$.
Other $E_{\text{MCA}}>0$, near $K$ and $H$, arises from $\langle d_{x^2-y^2/xy},\downarrow|L_z|d_{x^2-y^2/xy},\downarrow \rangle$.
In Fe$_{3}$GeTe$_{2}$, $d_{xz/yz}$ in $\uparrow$ channel are occupied near $KM$ and $H$, 
giving $E_{\text{MCA}}<0$ by $\langle d_{xz/yz},\uparrow|L_z|d_{xz/yz},\downarrow \rangle$ 
in spin-flip channel. 
Moreover, occupied $d_{x^2-y^2/xy}$ in $\uparrow$ channel near $K$ leads to $E_{\text{MCA}}<0$ through the spin-flip channel $\langle d_{x^2-y^2/xy},\uparrow|L_z|d_{x^2-y^2/xy},\downarrow \rangle$. As a result, in Fe$_{3}$GeTe$_{2}$, $|E_{\text{MCA}} [ \textrm{Fe(II)} ]|$ reduces.

So far, the $E_{\text{MCA}}$ difference between Fe$_{3}$GaTe$_{2}$ and Fe$_{3}$GeTe$_{2}$ has been investigated through Fe band analysis. 
Total $E_{\text{MCA}}$ of Fe$_{3}$GaTe$_{2}$ and Fe$_{3}$GeTe$_{2}$ are 1.38 meV/f.u. and 3.15 meV/f.u., respectively.
Among the two terms of the Hamiltonian mentioned earlier, the energy scale of MCA is smaller than that of $J$.
It insufficient to discuss $T_c$ with MCA alone.
Hence, the magnetic exchange coefficients are discussed by comparing  Fe$_{3}$GaTe$_{2}$ and Fe$_{3}$GeTe$_{2}$.

\begin{table*}[t]
\centering
\caption{
Magnetic exchange coefficients ($J_n$), effective exchange coefficients ($z_nJ_n$), and relevant atomic distances ($d_n$) of Fe$_{3}$GaTe$_{2}$ and Fe$_{3}$GeTe$_{2}$.
$J_1$ to $J_4$ (in meV) as sketched in Fig.\ref{fig:structure};
$J_n$ and $z_{n}J_{n}$ in meV, where $z_n$ is number of neighbors;
$d_n$ is in \AA.
}
\begin{ruledtabular}
  \begin{tabular}{lcccccc}
  &  & $J_{1}$ & $J_{2}$ & $J_{3}$ & $J'_{3}$ & $J_{4}$ \\
\hline
  & Type & Fe(I)$-$Fe(I) & Fe(I)$-$Fe(II) & Fe(I)$-$Fe(I) & Fe(II)$-$Fe(II) & Fe(I)$-$Fe(I) \\
\hline
  & $z_{n}$ & 1 & 6 & 12 & 6 & 6 \\
\hline
\multirow{ 3}{*}{Fe$_{3}$GaTe$_{2}$} & $J_{n}$ & 48.67 & 18.54 & 4.23 & 0.26 & 1.85\\
& $z_{n}J_{n}$ & 48.67 & 111.24 & 50.76 & 1.56 & 11.10 \\
& $d_{n}$ & 2.41 & 2.62 & 4.03 & 4.03 & 4.70 \\
\hline
\multirow{ 3}{*}{Fe$_{3}$GeTe$_{2}$} & $J_{n}$ & 78.84 & 24.61 & $-$2.46 & $-$4.56 & $-$2.13 \\
& $z_{n}J_{n}$ & 78.84 & 147.66 & $-$29.52 & $-$27.36 & $-$12.78 \\
& $d_{n}$ & 2.46 & 2.63 & 4.02 & 4.02 & 4.71 \\
\end{tabular}
\end{ruledtabular}
\label{tab:table1}
\end{table*}

\begin{figure}[t]
    \centering
    \includegraphics[width=1\columnwidth]{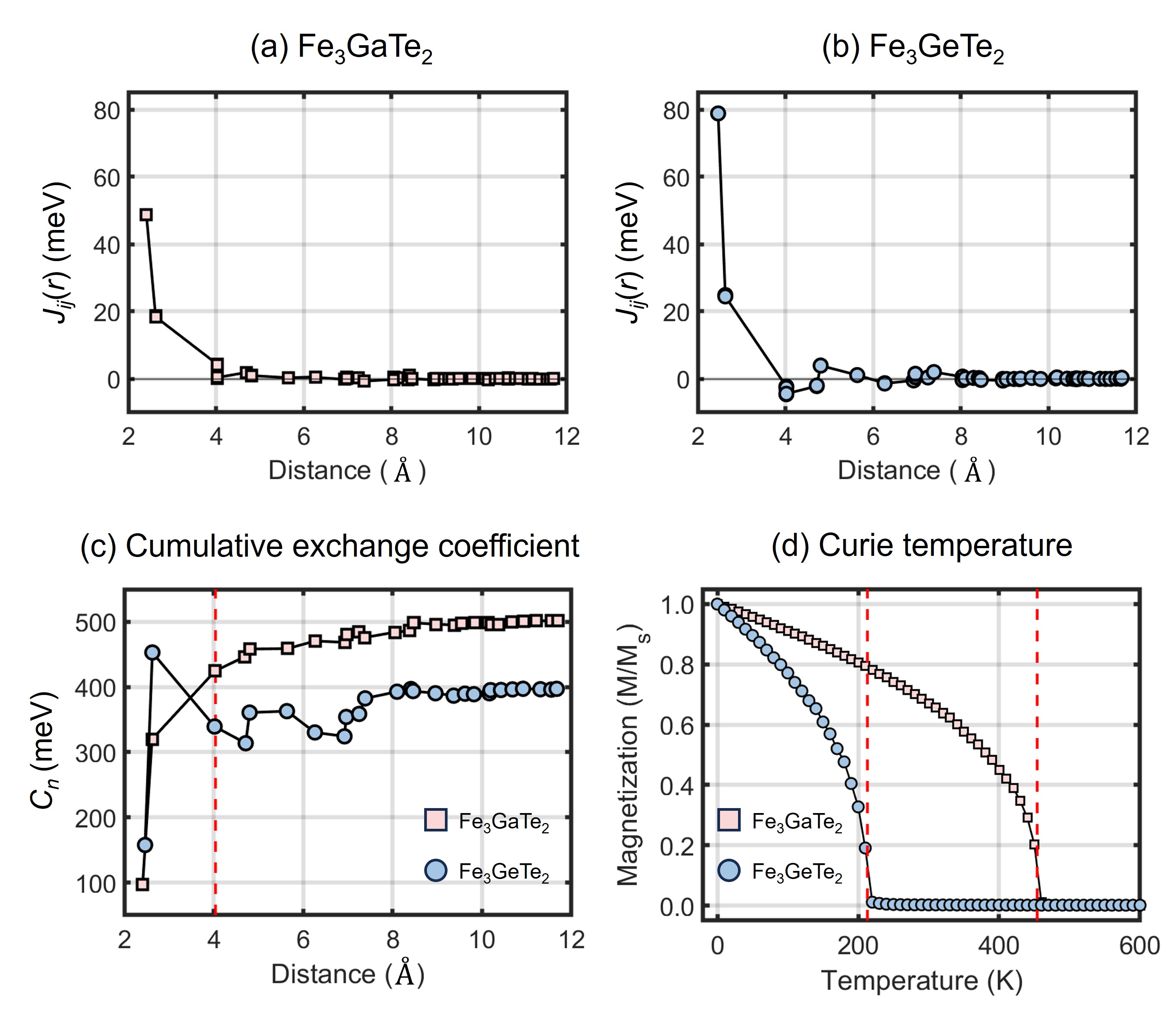}
    \caption{The exchange interaction coefficients as function of distance for (a) Fe$_{3}$GaTe$_{2}$ and (b) Fe$_{3}$GeTe$_{2}$.  (c) The cumulative all $J_{n}$, $C_n=\sum^n_i z_iJ_i$, 
    respect to the atomic distance. 
    (d) The normalized magnetization as function of temperature. Fe$_{3}$GaTe$_{2}$ and Fe$_{3}$GeTe$_{2}$ show $T_C$ of 456 K and 213 K, respectively. Red and blue spheres denote Fe$_{3}$GaTe$_{2}$ and Fe$_{3}$GeTe$_{2}$, respectively.}
    \label{fig:totalJ}
\end{figure}

\emph{Magnetic exchange coefficients and Curie temperature}$-$
Schematic of the magnetic exchange interactions is shown in Fig. \ref{fig:structure} (b),
where pair distance increases from $J_1$ to $J_4$.
$J_1$ represents the exchange interaction between vertical Fe(I) pairs, while $J_3$ and $J'_3$ correspond to horizontal interactions between Fe(I)-Fe(I) and Fe(II)-Fe(II), respectively.
$J_{2}$ and $J_{4}$ indicate diagonal interactions of Fe(I)-Fe(II) and Fe(I)-Fe(I), respectively.
Calculated $J_{ij}$ values under the distances
in Fe$_3$GaTe$_2$ and Fe$_3$GeTe$_2$ are shown in Fig. \ref{fig:totalJ} (a) and (b), respectively.
In our convention, positive (negative) value of $J$
reflects parallel (antiparallel) spin arrangement.

The nearest and second nearest neighbor, $J_{1}$ and $J_{2}$, are 48.67 meV (78.84 meV) and 18.54 meV (24.61 meV)
for Fe$_{3}$GaTe$_{2}$ (Fe$_{3}$GeTe$_{2}$), respectively.
As other studies have reported,
\cite{li2023tuning, li2023tremendous, ruiz2024origin}, 
$J_{1}$ and $J_{2}$ of Fe$_{3}$GaTe$_{2}$ are smaller than Fe$_{3}$GeTe$_{2}$.
In conventional wisdom, larger $J_{1}$ and $J_{2}$ imply higher $T_C$.
However, this is not in our case as $T_C$ of Fe$_{3}$GaTe$_{2}$ is more than 100 K higher than Fe$_{3}$GeTe$_{2}$ . 

In the mean-field approximation \cite{weiss1907hypothese, skomski2008simple}, $T_C$ for one magnetic sublattice is 
$T_{C}\approx\frac{1}{k_{B}}zJ_{1}$, where $z$ is the number of neighbors; $k_{B}$ is the Boltzmann constant. 
In this case, only one magnetic exchange coefficient, $J_{1}$ is necessary. On the other hand, when the number of sublattice is more than one, it is not so simple. 
For an Ising magnets containing two magnetic sublattices $A$ and $B$, $T_C$ is obtained from eigenvalues of 2$\times$2 interaction matrix \cite{skomski2008simple, Smart, binek2003ising, coey1996rare},
\begin{equation}
\scalebox{.94}{$
    T_{C}=\frac{1}{2k_{B}}(z_{1}J_{AA}+z_{2}J_{BB})+\frac{1}{2k_{B}}\sqrt{(z_{1}J_{AA}-z_{2}J_{BB})^{2}+4(z_{3}J_{AB})^2},
    \label{eq:MFT}
$}
\end{equation}
where $J_{AA}$ and $J_{BB}$ are the intra-sublattice exchange constants; $J_{AB}$ for the inter-sublattice interactions. 
Clearly, instead of single $J_1$, three magnetic exchange coefficients are necessary.
For the system with three magnetic sublattices, it is necessary to diagonalize 3$\times$3 matrix. 
Extension of Eq. \ref{eq:MFT} becomes more complicated, 
as six $J$ terms are necessary and one needs to solve cubic equation \cite{SMJ}.
In Fe$_{3}$GaTe$_{2}$ or Fe$_{3}$GeTe$_{2}$, the number of sublattices is more than two. As discussed, higher order $J$ (i.e., $J_{3}$ and above) cannot be ignored.

Table \ref{tab:table1} lists magnetic exchange coefficients, $J_n$ ($n=1,2,3,4)$ 
with distances between atoms ($d_n$).
As $J_n$ has more than one neighbor,
we introduce the effective magnetic exchange coefficients, $z_n J_n$,
where $z_n$ is number of neighbors. We note that in Eq. \ref{eq:MFT}, the effective magnetic exchange coefficients appear instead of the exchange coefficients ($J_n$). 
As listed in Table \ref{tab:table1}, while $J_1 > J_2$, with six neighbors ($z_2=6$), $z_2J_2 > z_1J_1$ for both Fe$_3$GaTe$_2$ and Fe$_3$GeTe$_2$.
Moreover, $z_3J_3$ is slightly larger than $z_1J_1$ in Fe$_3$GaTe$_2$. 
To easily visualize higher-order contribution, we further introduce the cumulative exchange coefficient,
$C_n = \Sigma^{n}_{i=1} z_iJ_i $, summation of the effective magnetic exchange coefficients.
Fig. \ref{fig:totalJ} (c) plots $C_n$ as a function of $d_n$. 
When $d_n<4~\AA$ ($n<3$), $C_n$ of Fe$_3$GaTe$_2$ is less than Fe$_3$GeTe$_2$.
When $d_n\geq4~\AA$ ($n\geq3$), as indicated by dashed line, Fe$_3$GaTe$_2$ has larger cumulative exchange coefficient than 
Fe$_3$GeTe$_2$ as $J_3,~J_3',~J_4 >0 $ for Fe$_3$GaTe$_2$ but $J_3,~J_3',~J_4 <0 $ for Fe$_3$GeTe$_2$.

\begin{figure}[t]
    \centering
    \includegraphics[width=1\columnwidth]{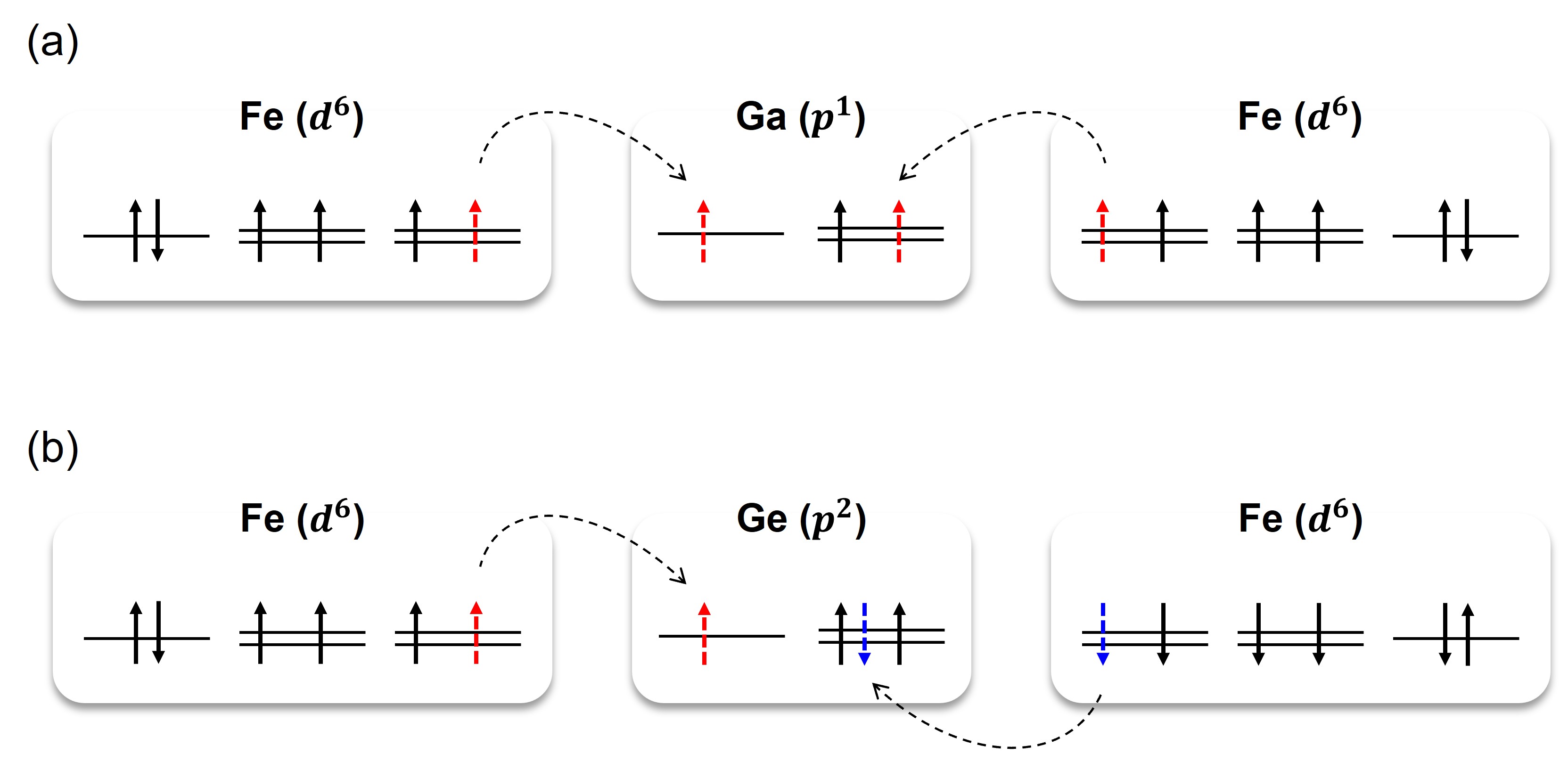}
    \caption{Schematics of the exchange interactions between Fe atoms for (a) Fe$_3$GaTe$_2$ and (b) Fe$_3$GeTe$_2$. Fe exchange interaction is mediated by hybridization with Ga or Ge. Nominal valency of Fe, Ga, Ge are $d^6$, $p^1$, and $p^2$, respectively.}
    \label{fig:diagram}
\end{figure}

The opposite signs of higher order $J$ are analyzed schematically in Fig. \ref{fig:diagram}. Fe-Fe exchange interaction is mediated by hybridization with Ga/Ge as Fe-Ga/Ge distance is much shorter than Fe-Fe one. We want to point out that this exchange by hybridization replaces hopping via ligand atom in conventional super-exchange model. Nominal valency of Fe is $d^6$ in good approximation; those of Ga and Ge are $p^1$ and $p^2$, respectively. Due to hexagonal symmetry, Fe $d$ orbital is split into $m=0, \pm1, \pm2$; Ga/Ge $p$ orbital into $m=0, \pm1$. For Fe$_3$GaTe$_2$, parallel spin configuration, $J_3$, $J_3'>0$, is possible by Fe-Ga $p$-$d$ hybridization. On the other hand, for Fe$_3$GaTe$_2$, Pauli exclusion principle prevents parallel spin configuration owing to $p^2$ occupation, hence $J_3$, $J_3'<0$. The hybridization of Fe-Ga/Ge is further analyzed and confirmed from partial density of states and orbital-resolved $J_3$ [See Supplementary Materials \cite{SMJ2}].

From Monte Carlo simulations using VAMPIRE software package \cite{metropolis1953equation, evans2014atomistic, vampire, CiteVampire}, the reduced magnetizations as a function of temperature are calculated for Fe$_3$GaTe$_2$ and Fe$_3$GeTe$_2$ as plotted in Fig. \ref{fig:totalJ} (d) with magnetic exchange coefficients up to 25th order, including the inter-layer interactions.
By fitting to $m(T)=(1-T/T_C)^\beta$ \cite{goldenfeld1992stat, evans2014atomistic}, 
where $m=M/M_s$ is the reduced magnetization for the saturation magnetization $M_s$; 
$\beta$ is the critical exponent.
From Fig. \ref{fig:totalJ} (d), $T_C$ are determined to be 456.5 K and 213.8 K 
for Fe$_3$GaTe$_2$ and Fe$_3$GeTe$_2$.
Our determined $T_C$ is higher (lower) than experiment Fe$_3$GaTe$_2$ (Fe$_3$GeTe$_2$).
Alternatively, $T_C$ are determined from eigenvalue equation in mean-field approximation for three magnetic sublattices,
including higher-order up to $J_6$ \cite{SMJ}.
By this, $T_C$ are 307.6 and 219.7 K for Fe$_3$GaTe$_2$ and Fe$_3$GeTe$_2$, respectively, 
much closer to experiment \cite{zhang2022above,deiseroth2006fe3gete2}.
The critical exponents, extracted from Fig. \ref{fig:totalJ}, are $\beta$=0.38 and 0.41 for Fe$_3$GaTe$_2$ and Fe$_3$GeTe$_2$, respectively, slightly larger than $\beta=0.325$ from three-dimensional Ising model \cite{Liu:2017srep}.
Furthermore, the scaling behavior of the magnetic exchange coefficients, 
$J(r)\approx 1/r^{\delta}$, with respect to distance are $\delta$=4.93 and 4.79 for Fe$_3$GaTe$_2$ and Fe$_3$GeTe$_2$, respectively, slightly larger than $\delta$=4.5 of mean-field and 4.6 of previous study for Fe$_3$GeTe$_2$ \cite{Liu:2017srep}.


\emph{Summary}$-$
In summary, 
Fe$_3$GaTe$_2$'s higher $T_C$ is investigated. 
Fe$_3$GaTe$_2$ has smaller $E_{\text{MCA}}$ (1.38 meV/f.u.) than Fe$_3$GeTe$_2$ (3.15 meV/f.u.).
The difference of MCA is analyzed from band structure perspective, where $E_F$ shift associated with one electron difference is responsible.
As contribution of the magnetic exchange is more dominant in spin Hamiltonian, 
magnetic exchange coefficients are exhaustively investigated up to sixth order or higher.
While Fe$_3$GaTe$_2$ has smaller magnetic exchange coefficients for the nearest and second nearest neighbors, $J_1$ and $J_2$, 
the cumulative exchange coefficients ($C_n=\sum^n_i z_iJ_i$), with fully accounting number of neighbors, 
are larger, which supports higher $T_C$ of Fe$_3$GaTe$_2$.
Furthermore, $T_C$ are determined as 456.5 and 213.8 K for Fe$_3$GaTe$_2$ and Fe$_3$GeTe$_2$, respectively,
using Monte Carlo simulation. Alternatively, $T_C$ from eigenvalue equation in mean-field approximation
are 307.6 and 219.7 K, respectively.

\begin{acknowledgments} 

This research is supported by National Research Foundation of Korea
(NRF-RS-2024-00407298, NRF-RS-2024-00449996 and NRF-2022M3H4A1A04096339).

\end{acknowledgments}

%

\end{document}